\documentclass
[preprint,twocolumn,prl,lengthcheck,showpacs,10pt,letterpaper]{revtex4}%
\usepackage{amsfonts}
\usepackage{amsmath}
\usepackage{amssymb}
\usepackage{graphicx}%
\providecommand{\U}[1]{\protect\rule{.1in}{.1in}}

\begin{document}
\preprint{ }
\title{Orientational Coupling Amplification in Ferroelectric Nematic Colloids}
\author{Fenghua Li$^{\text{1}}$, Oleksandr Buchnev$^{\text{2}}$, Chae Il
Cheon$^{\text{3}}$, Anatoliy Glushchenko$^{\text{4}}$, Victor
Reshetnyak$^{\text{5}}$, Yuri Reznikov$^{\text{2}}$, Timothy J.
Sluckin$^{\text{6}}$ and John L. West$^{\text{1}}$}
\affiliation{$^{\text{1}}$Liquid Crystal Institute, Kent State University, Kent, OH 44242,
USA }
\affiliation{$^{\text{2}}$Institute of Physics, Ukrainian Academy of Science, Kyiv, Ukraine }
\affiliation{$^{\text{3}}$Hoseo University, Baebang, Asan, Chungnam, 336-795, Korea }
\affiliation{$^{\text{4}}$University of Colorado at Colorado Springs, Colorado Springs, CO
80933, USA }
\affiliation{$^{\text{5}}$Kyiv Taras Shevchenko University, Kyiv, Ukraine }
\affiliation{$^{\text{6}}$School of Mathematics, University of Southampton, Southampton
SO17 1BJ, United Kingdom}

\begin{abstract}
We investigated the physical properties of low concentration ferroelectric
nematic colloids, using calorimetry, optical methods, infra-red spectroscopy
and capacitance studies. The resulting colloids normally remain homogeneous,
but the nematic orientational coupling is significantly amplified. In
particular cases, the nematic orientation coupling increases by 10\% for
particle concentrations of 0.2\%. A manifestation of the increased
orientational order is that the clearing temperature of a nematic colloid
increases up to 40 Celsius degrees compared to the pure LC\ host. A
theoretical model is proposed in which the ferroelectric particles induce
local dipoles whose effective interaction is proportional to the square of the
orientational order parameter.

\end{abstract}
\volumeyear{year}
\volumenumber{number}
\issuenumber{number}
\eid{identifier}
\date{\today}
\received[Received text]{date}

\accepted[Accepted text]{date}

\maketitle

Colloids in which the solute is liquid crystalline (LC) are known to possess
an extremely rich set of behaviors \cite{1,2,3,4,5}. The anchoring between the
LC and microcolloidal particles ($\mathrm{\geqslant}$\textrm{1}$%
\operatorname{\mu m}%
$) can produce long-range orientational distortions around the particles. This
results in strong inter-particle interactions -- sometimes repulsive and
sometimes attractive -- in the mesophase. The interactions can give rise to
well-ordered structures of particles in the liquid crystal matrix (both
lattices and chains) \cite{2,3}. However, in most cases a prerequisite for
interesting LC colloidal behavior has been a high concentration dispersion,
typically with particle volume fraction $\mathrm{c}_{part}\mathrm{\geqslant}%
$\textrm{30}\%. In such systems aggregated particles produce director
distortions extending over macroscopic scales. These suspensions scatter light
strongly, and possess unique structural, mechanical, electro- and
magneto-optical properties \cite{5,6}.

Recently, we have shown that even at low concentrations ($\mathrm{c}%
_{part}\mathrm{\leqslant}$\textrm{1}\%), LC colloids differ strongly from the
pure host material \cite{7,8,9,10}. These colloids consist of submicron
ferroelectric particles suspended in the LC host. In these systems, unlike in
classic LC colloids, the suspension-matrix interaction is insufficient to
disturb the LC orientation. This small concentration dramatically increases
the dielectric anisotropy, significantly decreases the Freedericksz transition
voltage, and significantly accelerates electric field-induced director reorientation.

In this letter, we report results which show that these phenomena are general
properties of liquid crystal suspensions containing ferroelectric colloidal
nanoparticles. The addition of impurities normally decreases the nematic
clearing temperature $T_{NI}$ \cite{11}. However, our measurements show
massive increases in $T_{NI}$, of the order of 40$%
\operatorname{{}^{\circ}{\rm C}}%
$, for mass impurity concentrations of the order of 0.2\%. These results imply
an increase in the effective nematic interaction parameter. Measurements of
the birefringence, dielectric anisotropy and order parameter in the suspension
are consistent with this picture.

We have also constructed a theoretical model. The ferroelectric particles in
the suspension produce large electric fields in their neighborhood. These
electric fields produce induced dipoles on the nematic molecules, and the mean
magnitude of the induced dipoles is proportional to the average polarizability
and hence the nematic order parameter. The dipolar interaction between the
induced dipoles is thus proportional to the nematic order parameter, and
provides an additional component to the effective nematic interaction. Order
of magnitude estimates of the magnitude of this quantity are consistent with
our observations.

Most of our experiments utilized the nematic LC MLC-6609 from Merck ($T_{NI}%
=$91.5$%
\operatorname{{}^{\circ}{\rm C}}%
$), doped with BaTiO$_{3}$ nanoparticles from Aldrich. The nanoparticle
dimension was 50--100\textrm{\ }$%
\operatorname{nm}%
$ (transmission electron microscopy). The spontaneous polarization of the
monodomain particles is assumed the same as that of a BaTiO$_{3}$ monocrystal,
0.26 $%
\operatorname{C}%
\operatorname{m}%
^{-2}$ \cite{12}. The detailed preparation process of the colloid was
described recently (see Ref. \cite{7}).

The clearing temperature $T_{NI}$ was determined using DSC at a rate of 5 $%
\operatorname{K}%
\operatorname{min}%
^{-1}$. The resulting DSC scans for the pure MLC-6609 and for the colloid
($\mathrm{c}_{part}\mathrm{\approx}$\textrm{0.2}\%) are shown in Fig. 1. When
the BaTiO$_{3}$ nanoparticles are introduced, the clearing temperature
increases by 38.7$%
\operatorname{{}^{\circ}{\rm C}}%
$. A lower mass fraction ($\mathrm{c}_{part}\mathrm{\approx}$\textrm{0.05}%
\%$)$ produced a reduced shift $\Delta T_{NI}=$23.1$%
\operatorname{{}^{\circ}{\rm C}}%
$. The changes in $T_{NI}$ were confirmed using polarized optical microscope.%

\begin{figure}
[tbh]
\begin{center}
\includegraphics[
height=2.3359in,
width=2.7077in
]%
{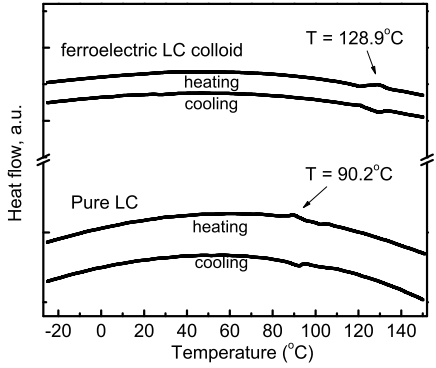}%
\caption{DSC-graphs for heating and cooling of the samples with pure LC
MLC-6609 (bottom pair of the curves) and colloids of BaTiO3 in MLC-6609 (top
pair of the curves), $\mathrm{c}_{part}\mathrm{\approx}$\textrm{0.2\%}.}%
\label{Fenghua_Graph1}%
\end{center}
\end{figure}

We also measured other LC properties in the colloidal suspension, in order to
check the hypothesis that the increase in $T_{NI}$ is a consequence solely of
an increased effective nematic interaction. Firstly we measured order
parameters, using FTIR spectroscopy and observing the dichroism of the
characteristic functional groups of the nematic matrix components. We used the
stretch vibration of C=C groups of benzene rings ($\mathrm{\nu=}$\textrm{1513
}$\mathrm{%
\operatorname{cm}%
}^{-1}$) because this group is oriented along the long axis of the LC
molecules. The C=C band dichroism was measured using a Magna 550 FTIR
(Nicolet) spectrometer in NaCl cells of $\mathrm{16}%
\operatorname{\mu m}%
$ thickness. The LC and the colloid were homogeneously aligned using a thin
($\mathrm{<}$\textrm{0.1}$%
\operatorname{\mu m}%
$) rubbed polyimide alignment layer. The order parameter was evaluated using
the formula $S=(D-1)/(D+2)$, where $D=A_{//}/A_{\perp}\ $is the dichroism of
absorbance parallel ($A_{//}$), and perpendicular ($A_{\perp}$) to the IR beam
polarization. Fig. 2 shows the temperature dependence of the resulting order
parameters, both in the pure MLC-6609 and in the ferroelectric colloid
($\mathrm{c}_{part}\mathrm{\approx}$\textrm{0.2}\%). The clearing temperature
increase is mirrored by an order parameter increase; at 30$%
\operatorname{{}^{\circ}{\rm C}}%
$ the order parameter gain, $S_{COL}/S_{LC}=$1.2.%

\begin{figure}
[tbh]
\begin{center}
\includegraphics[
trim=0.000000in 0.000000in -0.000794in 0.000000in,
height=2.3359in,
width=2.6826in
]%
{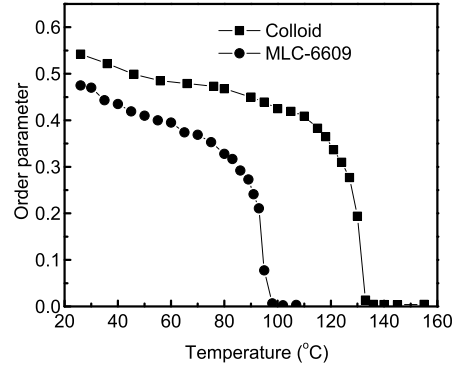}%
\caption{Temperature dependences of the order parameter of MLC-6609 and the
ferroelectric colloid ($\mathrm{c}_{part}\mathrm{\approx}$\textrm{0.2\%}).}%
\label{Fenghua_Graph2}%
\end{center}
\end{figure}

A further measurement compares the optical birefringence of the colloid to
that of the pure nematic. We assume the ferroelectric nanoparticles act as
effective molecular dopants. The birefringence of the colloid can then be
written as \cite{13}:%

\begin{equation}
n_{a}^{COL}=\frac{4\pi FN_{LC}\gamma_{a}^{COL}}{n_{e}^{COL}+n_{o}^{COL}%
}S_{COL} \label{eq1}%
\end{equation}
where $F$ is the local field factor, $\gamma_{a}^{COL}=(1-c_{part})\gamma
_{a}^{LC}+$\ $c_{part}(\gamma_{a}^{part})$ with $\gamma_{a}^{LC}$\ is the LC
molecular polarizability anisotropy and $\gamma_{a}^{part}$\ is the
polarizability anisotropy of the ferroelectric particles. The birefringence
gain with respect to the pure LC is then given by:%

\begin{equation}
\frac{n_{a}^{COL}}{n_{a}^{LC}}=\frac{(n_{e}^{LC}+n_{o}^{LC})}{(n_{e}%
^{COL}+n_{o}^{COL})}\frac{\gamma_{a}^{COL}}{\gamma_{a}^{LC}}\frac{S_{COL}%
}{S_{LC}} \label{eq2}%
\end{equation}

The birefringence gain in the colloid involves three separate components. The
first is the statistical mechanics-induced order parameter increase. The
second depends also on the effective anisotropy of both the ferroelectric
particles and LC molecules, the values of which are not known. Finally, the
gain depends on the ratio of the refractive indices of the LC and the colloid.
We measured these refractive indices and find that the ratio is essentially unity.

Experimentally we obtained $n_{a}(T)$ using an Abb\'{e} refractometer. An
independent set of experiments used a retardation technique measuring the
phase shift between e- and o-waves $\varphi=\pi dn_{a}/\lambda$\ in planar
cells ($d$ = 12 $%
\operatorname{\mu m}%
$). The results of the two sets of measurements were in agreement. In Fig. 3,
we show the temperature dependence of the birefringence in the pure LC and in
one colloidal sample ($c_{part}\approx$0.05\%). At 32$%
\operatorname{{}^{\circ}{\rm C}}%
$ the predicted $n_{a}^{COL}/n_{a}^{LC}=$1.2 agrees with the experimental
value $S_{COL}/S_{LC}=$1.2. Thus, the birefringence increase in the
ferroelectric colloid is consistent with the order parameter increase. We find
that at 32$%
\operatorname{{}^{\circ}{\rm C}}%
$, $\varepsilon_{a}^{COL}/\varepsilon_{a}^{LC}$= 1.54. This result is close
to, but not identical to $(n_{a}^{COL})^{2}/(n_{a}^{LC})^{2}\approx
S_{COL}^{2}/S_{LC}^{2}$= 1.44. We speculate that this discrepancy may be due
to an extra contribution from the ferroelectric colloidal particles to the
effective dielectric function of the colloid \cite{14}.%

\begin{figure}
[tbh]
\begin{center}
\includegraphics[
height=2.3359in,
width=2.9447in
]%
{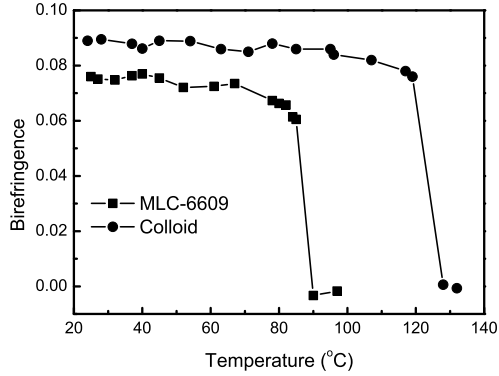}%
\caption{Temperature dependence of the birefringence of MLC-6609 and the
ferroelectric colloid ($\mathrm{c}_{part}\mathrm{\approx}$\textrm{0.05\%)}.}%
\label{Fenghua_Graph3}%
\end{center}
\end{figure}

We speculate that the birefringence gain is a consequence only of the increase
in the nematic ordering. If this is the case, then the ratio of the colloidal
low frequency (i.e. $\mathrm{\nu\approx}$\textrm{1} $\mathrm{%
\operatorname{kHz}%
}$) dielectric constant to that in the pure LC should be the same as that at
optical frequencies: $\varepsilon_{a}^{COL}/\varepsilon_{a}^{LC}=(n_{a}%
^{COL})^{2}/(n_{a}^{LC})^{2}\approx S_{COL}^{2}/S_{LC}^{2}$. We used a
Hewlett-Packard 4194A impedance analyzer to measure the LC cell capacitance.
The cells were 15 $%
\operatorname{\mu m}%
$ thick, and the dielectric anisotropy was determined by comparing results
with planar and homeotropic orientation. In Fig. 4, we present results for the
temperature dependence of the dielectric anisotropy $\varepsilon_{a}(T)$.%

\begin{figure}
[tbh]
\begin{center}
\includegraphics[
height=2.3359in,
width=2.8677in
]%
{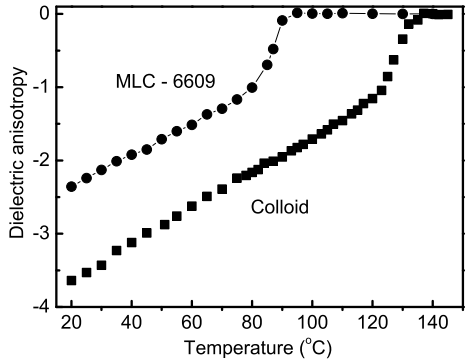}%
\caption{Temperature dependences of the dielectric anisotropy of MLC-6609 and
the ferroelectric colloid ($\mathrm{c}_{part}\mathrm{\approx}$\textrm{0.2\%}%
).}%
\label{Fenghua_Graph4}%
\end{center}
\end{figure}

Smaller increase in the clearing point have been observed for other LC host
systems (for example, Zli-2248, ZLI-4801 and 5CB), doped with both Sn$_{2}%
$P$_{2}$S$_{6}$ and BaTiO$_{3}$ nanoparticles. The increases are always
accompanied by increases in the dielectric anisotropy and birefringence.

All our experiments are consistent with an increase in the effective
orientational interaction, $\mathcal{U}$. According to the Maier-Saupe theory
\cite{15}, for our principal sample $\mathcal{U}_{COL}/\mathcal{U}_{LC}%
=T_{NI}^{COL}/T_{NI}^{LC}\approx$1.1. In order to make further progress, we
consider the additional terms in the free energy of the doped nematic LC
resulting from the presence of the colloidal particles. We assume for
simplicity that the ferroelectric particles are spherical in shape. A
ferroelectric particle possesses a large permanent polarization, $\mathbf{P}$,
which creates an electric field in its neighborhood \cite{16}:%

\begin{equation}
\mathbf{E}_{part}=\frac{R^{3}}{3\varepsilon_{0}}\left(  3\frac{\left(
\mathbf{P}\cdot\mathbf{r}\right)  \mathbf{r}}{r^{5}}-\frac{\mathbf{P}}{r^{3}%
}\right)  \label{eq3}%
\end{equation}
where $R$ is the particle radius. This electric field induces a dipole moment
in the host LC molecules, $\mathbf{\mu}_{ind}=\varepsilon_{0}\mathbf{\beta
}\cdot\mathbf{E}_{P}\sim1/r^{3}$, where $\mathbf{\beta}$ is the tensor of
molecular polarizability at low frequency. The interaction between the
electric field $\mathbf{E}_{P}$ and the induced dipole $\mathbf{\mu}_{ind}$ in
LC molecules is given by $U_{LC-P}\sim\sum\varepsilon_{0}\mathbf{\beta}%
_{i}\cdot\mathbf{E}_{P}(\mathbf{r}_{i})\mathbf{E}_{P}(\mathbf{r}_{i})$. The
induced dipole moments in LC molecules also interact with each other. If
$\varepsilon_{0}\mathbf{\beta}_{i}\cdot\mathbf{E}_{P}(\mathbf{r}%
_{i})=\mathbf{G}_{i}$ and $\mathbf{r}_{i}-\mathbf{r}_{j}=\Delta\mathbf{r}%
_{i,j},$ this additional intermolecular interaction is given by:%

\begin{equation}
U_{COL}^{add}\sim\sum_{\substack{i,j,j\neq i,\\particles}}\frac{\mathbf{G}%
_{i}}{4\pi\varepsilon_{0}}\left[  \frac{\mathbf{G}_{j}}{|\Delta\mathbf{r}%
_{i,j}|^{3}}-3\frac{\left(  \mathbf{G}_{j}\cdot\Delta\mathbf{r}_{i,j}\right)
\Delta\mathbf{r}_{i,j}}{|\Delta\mathbf{r}_{i,j}|^{5}}\right]  \label{eq4}%
\end{equation}

We now suppose the most important contributions come from nearest neighbor
molecules. Averaging eq.(\ref{eq4}) over LC molecule orientations yields:%

\begin{equation}
U_{a}\sim\frac{1}{2}N\sum_{\substack{i,\\particles}}\frac{\left(
\varepsilon_{0}\mathbf{\beta}_{a}S_{COL}\right)  ^{2}\left(  \mathbf{E}%
_{P}(\mathbf{r}_{i})\right)  ^{2}}{4\pi\varepsilon_{0}L^{3}}\equiv-\frac{1}%
{2}eS_{COL}^{2} \label{eq5}%
\end{equation}
where $N$ is the number of nearest neighbor LC molecules, $\mathbf{\beta}_{a}$
is the anisotropy of LC molecule polarizability at low frequencies and $L$ is
the nearest-neighbor distance. The key parameter $\mathit{e}$ is the
proportionality coefficient characterizing the response of the intermolecular
interaction to the field due to the ferroelectric particle.

In the Maier-Saupe molecular field theory, the intermolecular interaction
energy takes the form $F_{1}=-\frac{1}{2}\mathcal{U}S^{2}$ \cite{17}. The
parameter $\mathcal{U}$ is related to the nematic-isotropic phase transition
temperature by the ratio, $k_{B}T_{NI}=0.22\mathcal{U}$. This corresponds to
intermolecular interaction per unit volume $\mathcal{U}N_{LC}\approx
$4.7$\times$10$^{7}%
\operatorname{J}%
\operatorname{m}%
^{-3}$, where $N_{LC}$ is the LC molecular concentration. Therefore, we derive
a result for the change in the transition temperature due to a change in the
molecular interaction parameter, $\Delta T_{NI}\approx e{\small \cdot}%
T_{NI}{\small \cdot}\left(  \mathcal{U}N_{LC}\right)  ^{-1}$. We now derive an
explicit formula for the quantity $e$. In eq. (\ref{eq5}) the main
contribution to the sum comes from those molecules surrounding a given
ferroelectric particle. We calculate the sum explicitly, concentrating on
these LC molecules, giving rise to the following formula:%

\begin{equation}
e\approx\frac{1}{27}N{\small \cdot}N_{LC}{\small \cdot}N_{part}\frac
{\mathbf{\beta}_{a}^{2}P^{2}}{\varepsilon_{0}}\left(  \frac{R}{L}\right)  ^{3}
\label{eq6}%
\end{equation}
where $N_{part}$\ is the volume concentration of particles. To make
quantitative estimates, we take $P=$0.26 $%
\operatorname{C}%
\operatorname{m}%
^{\text{-2}}$, $N_{part}=$7$\times$10$^{\text{19}}%
\operatorname{m}%
^{-\text{3}}$, $N_{LC}=$2$\times$10$^{\text{27}}$ $%
\operatorname{m}%
^{-\text{3}}$, $\ N=$6, $R=$50 $%
\operatorname{nm}%
$, and $L=$0.5 $%
\operatorname{nm}%
$. Using $\mathbf{\beta}_{a}N_{LC}$\ $\approx\varepsilon_{a}$ \cite{13} and
$\varepsilon_{a}\approx$1, we find $e\approx$0.9$\times$10$^{\text{7}}%
\operatorname{J}%
\operatorname{m}%
^{-\text{3}}$. The resulting shift $\Delta T_{NI}\approx$70$%
\operatorname{{}^{\circ}{\rm C}}%
$ is in qualitative agreement with our experimental observations. Our model
does not consider other possible contributions to the intermolecular
interaction. There may be image forces or quadrupole interactions that remain
important even above the Curie temperature. These may be significant as a
result of the high nanoparticle dielectric constant, causing $\Delta T_{NI}%
$\ to be underestimated.

The observed effect requires $e$ as large as possible. From eq.(\ref{eq6})
this is governed by the large ratio $\left(  R/L\right)  ^{3}\approx
$10$^{\text{6}}$. For molecular, as opposed to colloidal, dopants this ratio
would be of the order of unity, dramatically reducing the effect $e/a=\Delta
T_{NI}$ $\sim$10$^{-4}$ $%
\operatorname{{}^{\circ}{\rm C}}%
$. It is the supramolecular size of the impurity particles which causes the
huge effect. In principle, this is testable by varying the particle size. The
effect is proportional to the particle surface area, but should not grow
without limit. Large particles generate orientational defects in their
neighborhood, which destroy the colloidal homogeneity and lower orientational
order. We estimate the critical particle size $R_{max}=l\sim K/W$ ($K$ is a
Frank constant) \cite{2}. For typical values $K=$ 10$^{-11}$ $%
\operatorname{N}%
$ and $W=$ 10$^{-4}$ \textrm{$%
\operatorname{J}%
\operatorname{m}%
$}$^{-2}$, $R_{max}$ is predicted as \textrm{100 }$%
\operatorname{nm}%
$.

In sufficiently large colloidal particles, the ferroelectric material is
expected to form a polydomain structure, which will further reduce the
effective interaction below the ideal $R^{2}$ dependence. In small systems,
the ferroelectric mean field will be insufficient to maintain a ferroelectric
ground state. Ferroelectricity usually vanishes for particles of dimensions
smaller than $\mathrm{\sim}$\textrm{10} $%
\operatorname{nm}%
$. We therefore speculate the optimal size of the ferroelectric particles in
homogeneous LC colloids to be in the range \textrm{(10--100) }$%
\operatorname{nm}%
$.

In conclusion, our results clearly show the unique properties of ferroelectric
nanoparticles/LC colloids. Since the particle size is much larger than the LC
molecular dimension, the strong electric field from the particles influences a
large number of neighboring LC molecules ($\mathrm{\sim}$\textrm{10}$^{3}$).
The huge ferroelectric dipole moment of the nanoparticles produces a powerful
field inducing dipolar intermolecular interactions that compete with the
spontaneous intermolecular interaction. The additional interaction then leads
to a dramatic increase in the LC mean field interaction and thence to a giant
increase in the clearing temperature.This also causes the homogeneous
ferroelectric LC colloid to have a higher birefringence, dielectric anisotropy
and order parameter than the pure LC material. These results are not only true
in the model systems which we have investigated in detail, but are a general
feature of colloidal suspensions made from ferroelectric colloidal particles
dispersed in thermotropic liquid crystal matrices. As a result ferroelectric
nanoparticles/LC colloids offer a simple and effective means to control the
physical properties of liquid crystalline materials.

\acknowledgements This work was supported by: NSF through Grant No.
DMR-0508137; Hoseo University, through the academic research fund in 2004; the
National Academy of Science of Ukraine (project \textquotedblleft Composite
liquid crystal and polymer materials for information
technologies\textquotedblright); the Royal Society of London, through a joint
Anglo-Ukrainian cooperation grant; and a NATO Collaborative Linkage Grant.

\end{document}